\documentclass[twocolumn,aps,prb,floatsfix,showpacs]{revtex4}
\usepackage{graphics}
\usepackage{amssymb,amsmath}
\usepackage{array}
\usepackage{dcolumn}

\begin{document}
 
\title{Influence of the structural
modulations and the Chain-ladder interaction in the
$S\!r_{14-x}C\!a_{x}C\!u_{24}O_{41}$ compounds.}

 \author{Alain Gell\'e}
 \affiliation{Laboratoire de Physique Quantique, IRSAMC~-~CNRS~UMR~5626,
 Universit\'e Paul Sabatier, 118 route de Narbonne, F-31062 Toulouse
 Cedex 4, FRANCE}

 \author{Marie-Bernadette Lepetit}
 \affiliation{Laboratoire de Physique Quantique, IRSAMC~-~CNRS~UMR~5626,
 Universit\'e Paul Sabatier, 118 route de Narbonne, F-31062 Toulouse
 Cedex 4, FRANCE}
 \affiliation{Laboratoire CRISMAT, ENSICAEN~-~CNRS~UMR~6508, 6
 boulevard Mar\'echal Juin, F-14050 Caen Cedex, France}

\date{\today}
 
\begin{abstract}
We studied the effects of the incommensurate structural modulations on
the ladder subsystem of the $S\!r_{14-x}C\!a_{x}C\!u_{24}O_{41}$
family of compounds using ab-initio explicitly-correlated
calculations. From these calculations we derived $t-J$ model as a
function of the fourth crystallographic coordinate $\tau$ describing
the incommensurate modulations. It was found that in the highly
calcium-doped system, the on-site orbital energies are strongly
modulated along the ladder legs. On the contrary the two sites of the
ladder rungs are iso-energetic and the holes are thus expected to be
delocalized on the rungs. Chain-ladder interactions were also
evaluated and found to be very negligible. The ladder
superconductivity model for these systems is discussed in the light of
the present results.

\end{abstract}
                                                                                
\maketitle

\section{Introduction}
The $S\!r_{14-x}C\!a_{x}C\!u_{24}O_{41}$ family of compounds has
attracted a lot of attention in the last years due to the discovery of
a superconducting phase~\cite{ObsSupra} in these quasi-unidimensional
systems. The $S\!r_{14-x}C\!a_{x}C\!u_{24}O_{41}$ are layered
materials with alternated planes of two-leg spin ladders and planes
of spin chains~\cite{struct1}. The superconducting phase, observed in
the highly calcium-doped systems, at low temperature and under
pressure, is believed to be the realization of the predicted
superconductivity in two-leg ladders systems~\cite{TheoEchSupra}.
Indeed, it is currently assumed that the
$S\!r_{14-x}C\!a_{x}C\!u_{24}O_{41}$ ladder subsystem can be
represented by a two-leg ladder isotropic $t-J$ model with $J/t\sim
0.4$ and a hole doping ranging between $1$~\cite{XRay00} and
$2.8$~\cite{COpt97} holes per formula unit (f.u.), that is between
$0.07$ and $0.2$ hole per ladder site. Numerical as well as
slave-bosons analytical calculations~\cite{supra}, showed that in this
parameter range, the superconducting pairing fluctuations are
dominant.  Hole-pairs are predicted to be bounded on the rungs and the
superconductivity to be supported by pairs collective modes.

The $S\!r_{14-x}C\!a_{x}C\!u_{24}O_{41}$ compounds present structural
incommensurate modulations of both the chains and ladder
subsystems. It was shown~\cite{srca} using ab-initio calculations,
that, unlike what was usually assumed, the chain subsystem cannot be
considered as a quasi-homogeneous system for which the structural
modulations induce only small perturbative effects. In fact, the
modulations caused by the influence of the ladders on the chain
subsystem are responsible for the low-energy physics of the latter,
that is charge localization and spin arrangement. These ab-initio
calculations showed that the parameters of the second neighbor $t-J+V$
model are strongly modulated. The major effects are (i) on the on-site
magnetic orbital energies that fluctuate within a few electron-Volt
range, (ii) on the first neighbor effective exchange integrals that
range from ferromagnetic (as expected for such $90^\circ$
oxygen-bridged copper sites) to antiferromagnetic (in the calcium
highly-doped compounds).  Moreover the calcium iso-electronic
substitution, that acts as a chemical pressure, is responsible for
large variations of the $t-J$ model parameters. It results a quite
different chain ground-state for the undoped and highly-doped systems.
The undoped compound exhibits the formation of second-neighbor dimers
caused by the electron localization on sites with low orbital energy,
while the large $x$ compounds exhibit the formation of low spin
clusters with antiferromagnetic local ordering.

One can thus wonder whether the structural modulations induce as
important effects on the ladder electronic structure, as observed on
the chain one. This point is of crucial importance since the existence
of such an effect would strongly questioned the interpretation of the
observed superconductivity. Another point, often addressed in the
literature, is the importance of possible chain-ladder
coupling. Indeed, in the highly calcium-doped compound, some of the
chain oxygen atoms come close to the ladder coppers as apical atoms.
The aim of the present work is to provide answers to these questions,
by the means of ab-initio calculations. The method used is an embedded
fragment spectroscopy method that explicitly computes both the
correlation effects and the screening effects. A $t-J$ model,
explicitly dependent on the structural modulations, is derived from
the ab-initio results for the ladders.  The calculations are done on
both ladders and mixed, chain and ladder, fragments, in the 
low-temperature phases of both the undoped~\cite{struct} and $x=13.6$
calcium-doped~\cite{5K} compounds. It would have been of great
interest to also run calculations in the superconducting phase,
unfortunately structural data are not available under
pressure. Nevertheless we will discuss possible extrapolation of the
ambient pressure results.

The next section will be devoted to the computational details. Section
3 will discuss the $t-J$ model, section 4 will discuss the
chain-ladder interactions and finally the last section will be devoted
to discussions and conclusions.

\section{Method and computational and details}


It is well known that, in strongly correlated systems,  few electrons
(per unit cell) are responsible for their spectacular
properties. These electrons are usually unpaired and localized both
spatially and energetically near the Fermi level (from now on we will
refer to them as magnetic electrons).  

Consequently, the electronic wave function of these systems is
essentially multi-configurational, i.e. it cannot be described, even
in a qualitative way, by a single Slater determinant. The consequence
is that single-reference based methods such as Hartree-Fock plus
perturbation theory or even Density Functional Theory (DFT) fail to
correctly treat some characteristics of the electronic structure of
such systems~\cite{dft}. While DFT methods yield reasonable Fermi
surfaces such as in the doped superconductor copper oxides, the LDA
method fails to predict the correct insulating versus metallic
behavior in a lot of correlated systems or effective local
interactions such as the magnetic exchange constant $J$.  This is for
instance the case for the same copper oxides for which the Fermi
surfaces were correctly predicted. Indeed, the LDA estimate of the
magnetic exchange in the $L\!a_2C\!uO_4$ compound ranges for instance
from $600\,\rm meV$ to $800\,\rm meV$ according to the functional used
while the experimental measurements~\cite{la2cuo4} yield $135\pm 6\
\rm meV$ and the embedded cluster method used in the present work
$145\ \rm meV$~\cite{DDCIhtc}.  LDA+U~\cite{LDA+U} or constrained
LDA~\cite{CLDA} (CLDA) methods that treat the correlation within the
magnetic orbitals in a mean field scheme yield better results. For
instance the LDA+U method yields between 129 and 165 meV for the $J$
value of $S\!r_2C\!uO_2C\!l_2$ (exp. 125 meV) and the CLDA method
yields 128 meV for the $L\!a_2C\!uO_4$ compound. These methods however
still present a lot of arbitrariness, for instance the uncertainty on
the results for the CLDA are of $50\%$ according to the authors of
ref.~\cite{CLDA}.

A few density functional calculations are present in the literature
for the present system. Arai and coworkers~\cite{DFT1} performed a LDA
calculation on the average structure (without the subsystem
modulations) of the $S\!r_{14}C\!u_{24}O_{41}$ in quasi periodic
approximation. Their results exhibit a metallic behavior for both the
chains and ladders subsystems whereas they are experimentally
found to be insulating. More recently, unpublished results from
Schuster and Schwingenschlögl~\cite{DFT2} found the same results when
including the structural modulations. Let us point out that the LDA+U
method should yield better results on such strongly correlated
systems. Unfortunately calculations  including
the structural modulation necessitate a much too large quasi-periodic
unit cell for the calculation to be tractable.

For such problems it is therefore necessary to rely on wave-function,
multi-reference, ab-initio, spectroscopy methods such as the
Difference Dedicated Configurations Interaction~\cite{DDCI,LMt} (DDCI) on
embedded clusters. This method proved its ability to quantitatively
reproduce the local interactions of a large family of magnetic systems
such as high $T_c$ copper oxides~\cite{DDCIhtc}, vanadium
oxides~\cite{vana}, nickel and cuprate fluorides~\cite{DDCI_Ni}, etc.
Such multi-reference wave function methods are very powerful to treat
strong correlation problems, however they necessitate a formally
finite system. The use of such methods for the treatment of strongly
correlated materials is possible since the interactions between the
magnetic electrons are essentially local. In fact it was proved, both
by quasi-degenerate perturbation analysis~\cite{revue} and numerically
by increase of the cluster size~\cite{XJloc,JP_Cu4,sr14,ca13} that only the
bridging ligands and the first coordination shell of the magnetic
centers are of importance for the magnetic, transfer or coulomb
interactions. 

The DDCI method is a variational method specifically designed for the
calculation of excited states involving open shells.  It is based on
the exact diagonalization of a selected configuration space built as
follow. First a partition of the cluster orbitals in occupied, active
and virtual is defined. The active set of orbitals is defined to be
the magnetic orbitals supporting the low energy physics as defined
above. In the present work this set of orbitals is formed by the open
$3d$ orbitals of the copper atoms, that is the $3d_{x^2-z^2}$ orbitals
for the ladder subsystem and the $3d_{xz}$ orbitals for the chain
subsystem ($x$ refers to the crystallographic ${\bf a}$ direction and
$z$ to the ${\bf c}$ direction). The occupied orbitals are the inner
orbitals that one can expect to be essentially doubly-occupied in the
system and the virtual orbitals are the orbitals essentially empty in
ground and low energy states of the system. The set of all
determinants that can be built from all possible spin and charge
arrangements within the active orbitals, all occupied orbitals being
doubly-occupied and all virtual orbitals being empty, is called the
Complete Active Space (CAS) and is considered as the reference space.
The exact diagonalization within this space will properly treat both
the open-shell character of the problem and the electronic correlation
within the active or magnetic orbitals. The simultaneous optimization
of the orbitals will yield the best possible partition (CASSCF
procedure). Once the reference space is defined the DDCI space to be
diagonalized is built from the CAS determinants and all single and
double excitations on each of them except the ones that excite two
occupied orbitals toward two virtual ones. Indeed, one can see easily
from a perturbative analysis that at the second order the later
configurations bring only a diagonal shift and do not contribute to
the excitation energies between the low energy states that essentially
differ by their component on the CAS.

The DDCI method has proved to be   very reliable
for the study of the strongly correlated compounds, since, the use of
a CAS ensures to treat properly~:
\begin{enumerate}
\item the multi-determinantal and open-shell character of the reference
wave function,
\item the correlation between the magnetic electrons, 
\end{enumerate}
while the selected doubly excited configurations allows to
include~\cite{LMt}
\begin{enumerate}
\item the screening effects (dynamical polarization and correlation)
on the different configurations of the reference wave function,
\item the effects of the ligand to metal charge transfers mediating
the interactions,
\item and the screening effects on the latter configurations.
\end{enumerate}
One could be surprised by the importance we give to the ligand to
metal charge transfer configurations, however their importance can be
shown by a simple perturbative analysis of the microscopic mechanism
supporting the effective exchange and transfer integrals.
\begin{figure}[h] 
\resizebox{!}{1.2cm}{\includegraphics{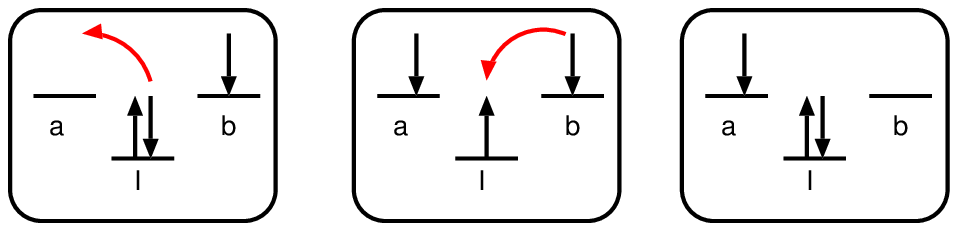}} \\[1ex]
\resizebox{!}{1.2cm}{\includegraphics{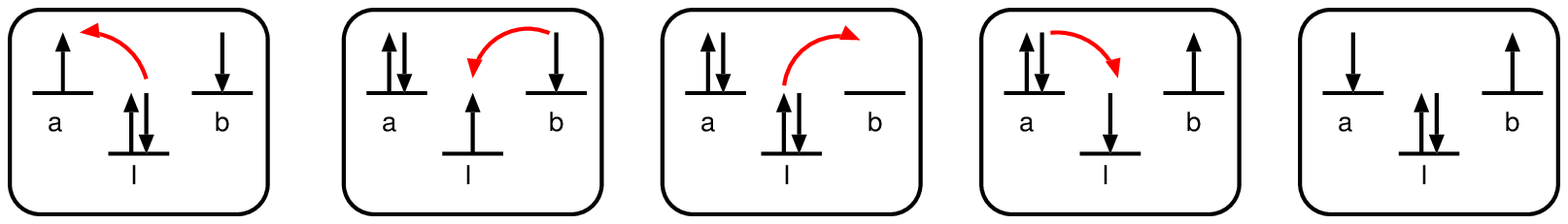}}
\caption{Schematic representation of the microscopic mechanism
responsible for the through-bridge transfer (first line) and
super-exchange (second line) effective integrals between the a and b
magnetic orbitals. The arrows picture the microscopic hopping involved
at each step between the magnetic a or b orbitals and the ligand one,
l. }
\label{f:tj}
\end{figure}
This mechanisms lead to the classical formulation 
\begin{eqnarray} \label{eq:tj}
t& \simeq &  \frac{t_{al}t_{bl}}{\Delta^\prime} \nonumber \\
J& \simeq & - 4\frac{t_{al}^2t_{bl}^2}{\Delta^2 U}
\end{eqnarray}
where $\Delta^\prime$ and $\Delta$ are the ligand to metal excitation
energies (let us note that these two energies are somewhat different
due to the presence or absence of a spectator electron on the
metal-ligand bond), $U$ is the Coulomb repulsion in the magnetic
orbitals and $t_{al}$, $t_{bl}$ are the ligand to metal direct
transfer integrals.  One sees from figure~\ref{f:tj} and
equation~\ref{eq:tj} that in order to obtain an accurate determination
of the effective transfer and exchange integrals, one should properly
evaluate the ligand to metal excitation energy. 

The effects of the rest of the crystal on the computed fragments are
essentially the Madelung potential and the exclusion effects.
Exclusion effects are treated using total-ion
pseudo-potentials~\cite{TIPS} (TIPs) that prevent the quantum fragment
electrons to extend in regions of space normally occupied by the
electrons of the rest of the crystal.  In the present work
TIPs~\cite{pseudo} are used on the first shell of atoms surrounding
the computed cluster. 
The Madelung potential is reproduced by a set of point
charges located on the material atomic position. The charges were
chosen so that the strontium and calcium atoms are di-cations
($C\!a^{2+ }$ and $S\!r^{2+}$). The copper and oxygen charges where
chosen as $C\!u^{2+}$ and $O^{-2+\delta}$.  The $\delta$ correction is
associated to the self-doping in holes. It was attributed to the
oxygens since the holes were shown to be essentially localized on
them~\cite{sr14}. The $\delta$ values for the chain and ladder
subsystems were chosen in agreement with the experimental assumptions
and our previous calculations~\cite{sr14,ca13}, that is all the holes
on the chains for the $S\!r_{14}C\!u_{24}O_{41}$ system and one hole
transfer on the ladders for the $S\!r_{0.4}C\!a_{13.6}C\!u_{24}O_{41}$
compound.  Concerning now the hole repartition within each subsystem,
we have chosen an homogeneous repartition, corresponding to an average
state in order not to induce localization effects due to a particular
repartition.  One should note that the main effect of a localization
of the magnetic electrons will be a modification of the on-site
orbitals energies through the Coulomb repulsion. This self-consistent
effect is taken into account in this work through the inclusion of the
screened NN electron repulsion in the energy curves (see
figure~\ref{f:ece}). Due to screening effects, larger distance coulomb
repulsion is expected to be of much weaker importance.

\begin{figure}[ht] 
\resizebox{8cm}{!}{\includegraphics{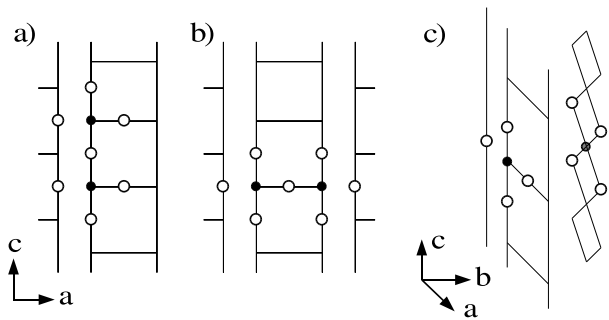}}
\caption{Schematic representation of the computed fragments. a) for
the ladder legs interactions, b) for the ladder rungs interactions and
c) for the chain-ladder interactions.  Circles represent the fragment
atoms~; the gray circles represent oxygen atoms, while the black
circles represent the copper atoms.}
\label{f:frag}
\end{figure}

Two types of fragments are used in the present work~:
\begin{itemize} 
\item $C\!u_2O_7$ fragments for the ladder legs and the 
ladder rungs interactions (see figures~\ref{f:frag}a and ~\ref{f:frag}b), 
\item $C\!u_2O_{8}$ for the chain-ladder interactions (see
figure~\ref{f:frag}c).
\end{itemize}

On-site orbital energies ($\varepsilon$) and hopping ($t$) parameters
of the $t-J$ model are extracted from the first doublet-doublet
excitation energies, and associated wave-functions, of the above
fragments with one magnetic electron. Exchange integrals ($J$) are deduced
from the singlet-triplet first excitation energies on the same fragments 
with two magnetic electrons.  A least-square fit method is used in
order to fit the effective parameters to the ab-initio results.

The calculations were done using the MOLCAS~\cite{molcas} and CASDI
set of programs and the basis sets used can be found in
reference~\cite{bases}.

In an incommensurate system all atoms or dimers have different
geometries and environment. It is therefore impossible to compute
all symmetry independent set of $t$, $J$, $\varepsilon$, etc. parameters.  
Indeed, the atomic positions
of the present compounds are described as the three dimensional projection
of a periodic four dimensional structure (for further details see
reference~\cite{struct4}). It comes for the atomic positions of the two
subsystems  
$$ \vec r_i(p,n_x,n_y,n_z) = \vec r_i(p,0,0,0) + \vec
R_i(n_x,n_y,n_z) + \vec f\left(\tau_i(n_z)\right)$$ where $i$ refers
to the subsystem and $p$ labels the atoms in the unit cell,
$$\vec r_i(p,0,0,0) = x_0(p) \vec a + y_0(p)\vec b + z_0(p)\vec c_i$$
is the atomic position of the average structure in the reference unit
cell 
$$\vec R_i(n_x,n_y,n_z) = n_x \vec a + n_y \vec b + n_z \vec c_i$$
is the translation vector of the average structure of each subsystem
$i$, and $\vec f\left(\tau_i(p,n_z)\right)$ is the modulation vector of the
subsystem $i$ with the periodicity of the other subsystem~: $j$.
$$\tau_i(p,n_z) =
(z_0(p) + n_z){ c_i \over c_j}$$ and  
\begin{eqnarray*}
\vec f\left(\tau_i(p,n_z)\right) &=& 
\sum_{n=1}^{n_{max}} A_{i,n}(p) \sin\left(2\pi n
\tau_i(p,n_z)\right) + \\ &&B_{i,n}(p) \cos\left(2\pi n
\tau_i(p,n_z)\right)
\end{eqnarray*}

We thus computed the $t-J$ model parameters at 11 different positions
along the ladders.  These positions were chosen in order to get a good
representation of the incommensurate distortions. The computed values
were then extrapolated to the whole system using the above crystallographic
description of incommensurate systems. It was done using a Fourier's
series analysis, as a function of the fourth coordinate $\tau$,
associated with the incommensurate modulations~\cite{struct4}. All
results of this work are given as a function of the $\tau$ coordinate
of the copper atom (on-site parameters) and of the copper atom of
lowest $x$ and $z$ coordinate for the interaction terms.

\section{Influence of the structural modulations on the ladder subsystem}

\subsection{$t-J+V$ model }

\begin{figure}[ht] 
\resizebox{8.5cm}{!}{\includegraphics*{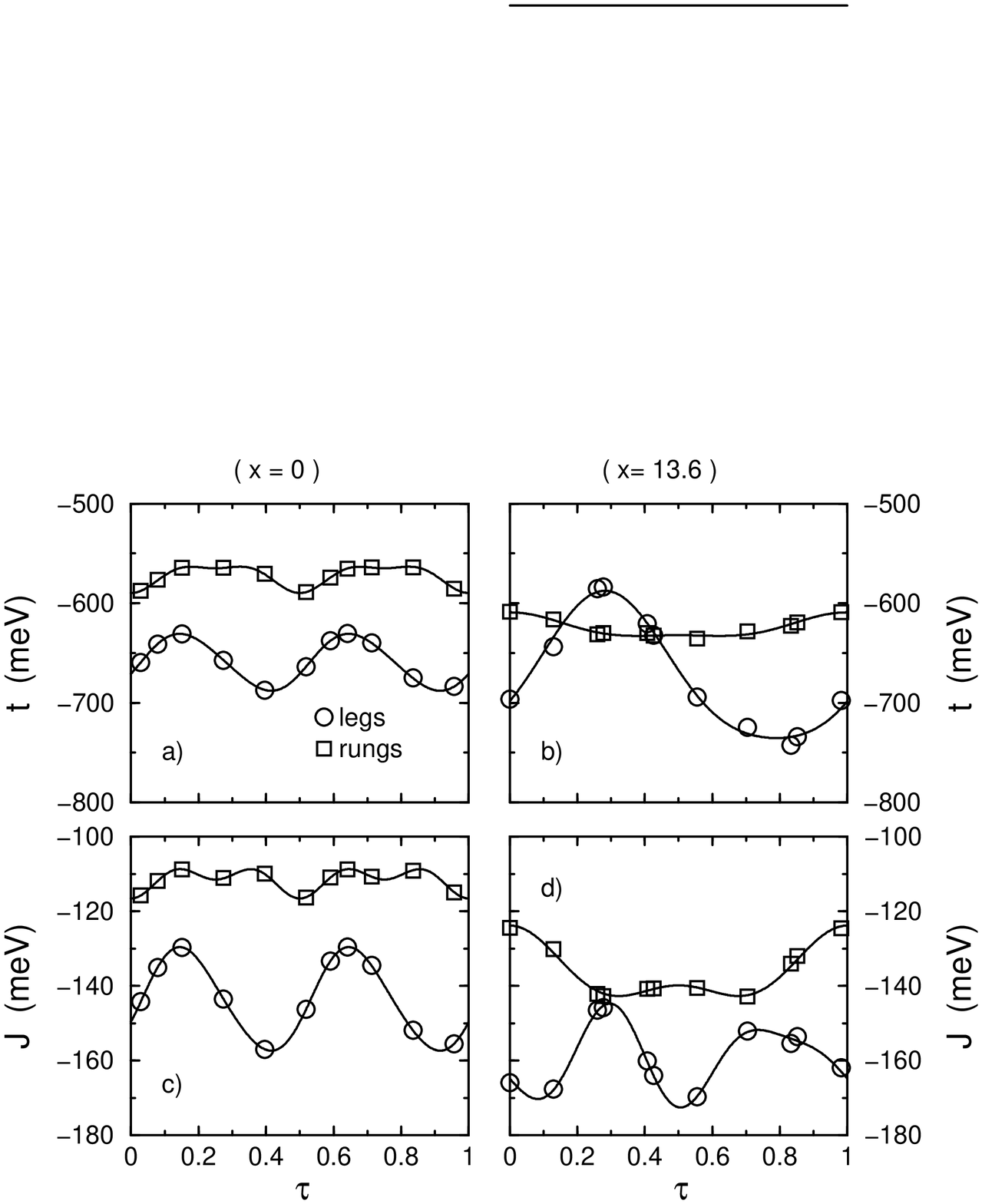}}
\caption{Effective hopping (a and b) and exchange (c and d) parameters
as a function of the modulation crystallographic coordinate
$\tau$. Figures a and c correspond to the undoped compound and figures
b and d to the calcium-doped compound ($x=13.6$). The circles and
squares are the computed points for the legs and rungs, the solid line
are the fitted Fourier series.}
\label{f:tJ}
\end{figure}

Figure~\ref{f:tJ} displays the hopping and exchange parameters for the
undoped and calcium-doped compounds. One first notices that, as was
observed on the chains subsystem~\cite{sr14,ca13}, the variations of
the parameters are quite large. For instance, on the $x=13.6$ ladder
legs, the hopping modulations reach up to $160\, \rm meV$.  One should
however point out that the nominal values of both the hopping and
exchange integrals are much larger for the ladder subsystem than for
the chain one due to the nearly $180^\circ$ $C\!u$--$O$--$C\!u$
angles. In fact, this is essentially the ligand to metal charge
transfer integrals of equations~\ref{eq:tj} that are sensitive to the
structural modulations. In the chain subsystem one of them, let us say
$t_{al}$, is nearly zero (due to the nearly 90$^circ$ angles) and thus
very responsive to any structural distortions. In the ladder
subsystem both $t_{al}$ and $t_{bl}$ are large (due to the nearly
180$^circ$ angles).  The consequence is that the relative variations
of the parameters are much smaller on the ladder subsystems. Indeed,
even for the hopping on the ladder legs of the $C\!a_{13.6}$ compound,
the relative variation reaches only $22\%$ of the nominal value, while
it was $100\%$ for the chains nearest-neighbor (NN) hoppings. It can
thus be expected that despite their large absolute values, hopping and
exchange fluctuations will not be as crucial for the ladders
electronic structure, as for the chains subsystems one. Nevertheless
one should remember that these modulations are not small enough for
their effect to be negligible.

Eccleston {\it et al}~\cite{Neut98} reported exchange values extracted
from neutron scattering experiments, for the undoped compound. These
values of $130\, \rm meV$ for the ladder legs and $72\, \rm meV$ for
the ladder rungs are to be compared with our computed mean values of
respectively $143\, \rm meV$ and $111\, \rm meV$. While the order of
magnitude are in good agreement the ratio between the average rungs
and legs exchange is larger in our calculations ($J_\bot/J_\| = 0.77$)
than in the experimental evaluation ($0.55$). One should however
notice that the $J_\bot/J_\|$ evaluation from neutron scattering does
not take into account the four copper cyclic exchange
parameter. Matsuda {\it et al}~\cite{cycl} found that the
consideration of this cyclic term in the fit of the neutron scattering
data of the $L\!a_6 C\!a_8 C\!u_{24}O_{41}$ compound notably increases
the a $J_\bot/J_\|$ ratio, more in agreement with the present results.
In addition, the cyclic exchange has been evaluated by ab-initio
methods for the $S\!r C\!u_2 O_3$ compound~\cite{SrCu2O3} and found to
be $18\%$ of the leg exchange value, in qualitative agreement with the
neutrons scattering data of Matsuda {\it et al} which evaluate this
term to $15\%$ of the leg exchange. Another argument in favor of the
quality of our calculations is the consistency between the exchange
and hopping parameters, computed independently. Indeed,
equation~\ref{eq:tj} yields $ J_\bot/J_\| \simeq
\left(t_\bot/t_\|\right)^2$, while our computed average values gives 
$J_\bot/J_\| = 0.77$ and $\left(t_\bot/t_\|\right)^2 = 0.75$.

It is interesting to notice that our calculation shows a variation of
the exchange constant along the ladder legs of about $19\%$ of the
exchange average value. This result compares nicely with the
predictions of Schmidt {\it et al}~\cite{kai} that postulated that the
structural modulations in the ladders would induce a transfer integral
modulation and thus an exchange integral one. In addition, they showed
that such a modulation ($\sim 16\%$ in their work) of the leg exchange
values was able to explain the discrepancies observed between the
Raman spectra of the $S\!r_{14} C\!u_{24} O_{41}$ and the $L\!a_6
C\!a_8 C\!u_{24} O_{41}$ compounds.

One can point out that both hopping and exchange absolute values
increase with the calcium content. This is specially true for the
ladder rung exchange that goes from a mean value of $112\ \rm meV$ for
the $x=0$ compound to $136 \ \rm meV$ for the $x=13.6$ compound. 

\begin{figure}[ht] 
\resizebox{8.5cm}{!}{\includegraphics{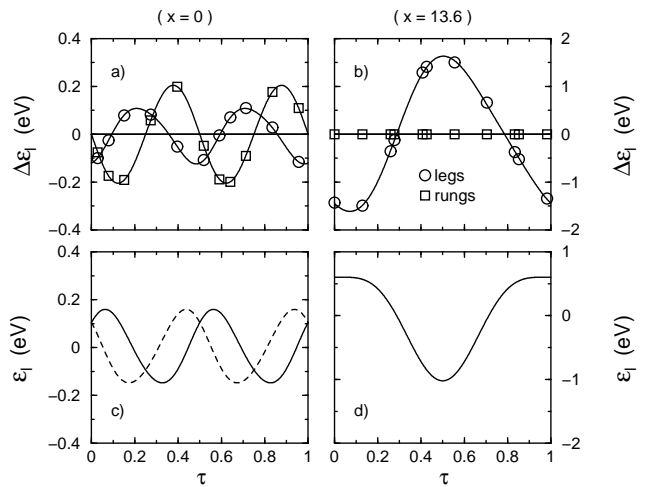}}
\caption{On-site orbital energies differences between NN sites (a for
 $x=0$ and b for $x=13.6$). The circles correspond to the computed
 values on the legs fragments and the squares to the rungs ones. The
 lines are the fitted Fourier series.  Deduced on-site energies are
 presented on figures c for $x=0$ and d for $x=13.6$. The solid and
 dashed curves correspond to the two legs of a ladder.}
\label{f:de}
\end{figure}

\begin{figure}[ht] 
\resizebox{6cm}{!}{\includegraphics{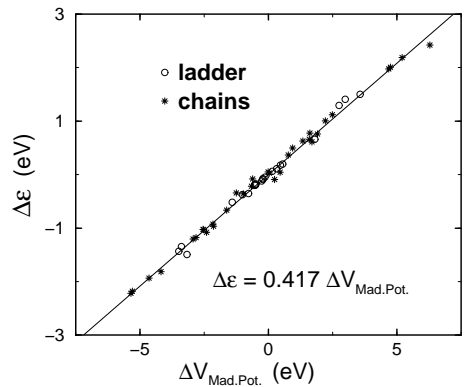}}
\caption{Computed NN-difference of the on-site orbital energies as a
function of the corresponding Madelung potential differences.}
\label{f:emad}
\end{figure}

Figure~\ref{f:de} reports the on-site orbital
energies, $\varepsilon(\tau)$, for the $x=0$ (c) and $x=13.6$ (d)
compounds. One sees immediately that the variations of these on-site
energies are large, however much weaker in amplitude than for the
chain subsystems. Indeed, for the undoped compound, $\varepsilon$
varies in a $0.3\ \rm eV$ range while it is $1.2\ \rm eV$~\cite{sr14}
for the chain subsystem. Similarly for the $x=13.6$ compound the
variations range is $1.6\ \rm eV$ for the ladders while it is $2.4\
\rm eV$~\cite{ca13} for the chains. This results are in agreement with
the hopping and exchange results and are due to the facts that (i) the
structural modulations are of weaker amplitude on the ladder
subsystem than on the chain one, (ii) the parameters modulation of
one subsystem, and specifically the on-site orbital energies, are
essentially due to the structural modulations of the concerned subsystem
itself~\cite{sr14}. It is remarkable that these on-site energies
modulations are totally correlated with the Madelung potential
modulation, both for the chain and ladder subsystems and both for the
undoped and highly calcium doped compounds (see
figure~\ref{f:emad}). Indeed, despite the fact that the local
environments of the ladders and chain magnetic centers are very
different the computed on-site orbital energies exhibit the same
dependence to the Madelung potential~: $\Delta\varepsilon = 0.42
\Delta V_{Mad.Pot.}$ (where $V_{Mad.Pot.}$ is the Madelung potential
on the copper sites and delta refers to the NN differences).

Comparing the $\varepsilon$ values as a function of the calcium
doping, one sees that as for the chain subsystem the calcium-doped
compound presents much larger on site orbital energy fluctuations
than the undoped system. This difference is in fact much larger than
for the chain subsystem since the $x=13.6$ versus $x=0$  average
orbital energy ratio is $5.3$ for the ladders while it is only $2$ for
the chains. Another important point, that will be discussed in more
details later on, is the fact that the $\varepsilon$ variation range is
weaker than the hopping integral in the $x=0$ system ($0.3\ \rm eV$
versus $0.660\ \rm eV$), while it is much larger in the $x=13.6$
compound ($1.6\ \rm eV$ versus $0.670\ \rm eV$) and can be expected to
induce in this case a hole localization on the ladders. 

Another important remark concern the orbital energy difference between
two sites belonging to the same rung. Indeed, for the undoped system
this difference is strongly modulated, with a NN energy difference
larger than between NN leg atoms (see figure~\ref{f:de}a).  For the
highly doped compound however, the two copper sites of a same rung are
iso-energetic. Indeed, the maximum absolute value of the computed
orbital energy difference between two sites on the same rung is
smaller than $1\ \rm meV$.  This energy degeneracy witnesses a hidden
extra symmetry since the two copper sites of a rung respectively
correspond to $\tau$ and $-\tau$. One retrieves this symmetry on
figure~\ref{f:de}d where the $\epsilon(\tau)$ function is symmetric
around $\tau=0.5$. The expected consequence of such a symmetry is the
fact that the holes present in the ladder subsystem should be
delocalized between the two sites of the ladder rungs, while localized
in the $\vec c$ direction.

\section{Chain-ladder interactions}

In this section we will address the possibility of a chain-ladder
coupling due to the structural modulations. Indeed, as mentioned in
the introduction, in the highly calcium-doped compounds, the chains 
can be distorted in such a way that their oxygen atoms  come into an
apical position of the ladder copper atoms with a small
$C\!u_{ladder}$--$O_{chain}$ distance ($2.7$ \AA). In the undoped
compound however, the $C\!u_{ladder}$--$O_{chain}$ distances
always remain larger than $3.1$ \AA.

\begin{figure}[ht] 
\resizebox{8.5cm}{!}{\includegraphics{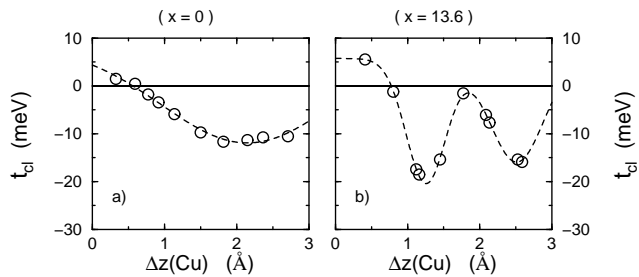}}
\caption{Effective hopping between the chain and ladder closest copper
sites as a function of the difference between the two copper
coordinates along the $\vec c$ axis ($\Delta z(C\!u)$). Circles
represent the computed points and the dashed lines are a guide for the
eyes.}
\label{f:ech-ch}
\end{figure}

Figure~\ref{f:ech-ch} displays the effective hopping integrals between
chain and ladder closest copper sites. One sees immediately that
unlike what has been supposed in the literature, the chain-ladder
hopping terms remain weak whatever the $C\!u_{ladder}$--$O_{chain}$
distances. Indeed, both the nominal values and the variation ranges
(in meV units~: $\left[-11.7,1.5\right]$ for the $x=0$ system and
$\left[-18.5,5.5\right]$ for the $x=13.6$ compound) are of the same
order of magnitude for the undoped and calcium-doped compounds. 

We also computed the chain-ladder magnetic couplings and the on-site
orbital energy differences between the chain and ladder systems. The
effective exchange were always found weaker than $1\ \rm meV$, that is
negligible. As far as the chain-ladder orbital energy differences are
concerned, the results obtained from the present chain-ladder
fragments are in full agreement with the results presented on the
ladder subsystems and those from references~\cite{sr14,ca13} on the
chain subsystems.  One should point out that the present calculations
allow us to set the chain-ladders relative energy levels, as will be
discussed further. Let us also notice that the orbital energy
differences between chain and ladder sites are not proportional to the
Madelung potential differences on the related sites, unlike what was
observed for the intra-subsystems orbital energy modulations (see
figure~\ref{f:emad}).

One should notice at this point, that counter ions were not included
in the fragment since their effects were expected to be smaller than
the contribution due the $C\!u$--$O$--$C\!u$ path. Their contributions
indeed involve $C\!u$--$O$--$C\!a/S\!r$--$O$--$C\!u$ path with
$C\!a/S\!r$--$O$ distances larger than $2.4$ \AA. As the obtained
values of the interactions are very small, their effects could be
not totally negligible. Nevertheless, if the inclusion of the counter
ions could slightly modify the values of the interactions, it will not
change the main conclusion which is that the ladder-chain interactions
are largely smaller than the intra-chain or intra-ladder interactions.
Both hopping and exchange values are found more than one order of
magnitude weaker. The chain and ladder layers can thus be safely
considered as non-interacting.

\section{Discussion and conclusion}

One of the most controversial subject about the
$S\!r_{14-x}C\!a_{x}C\!u_{24}O_{41}$ family of compounds is the chain
to ladder hole transfer as a function of the calcium content. While it
is clearly established that the hole transfer increases both with the
calcium content and the applied pressure, the actual amount of
transferred holes does not reach a consensus, even for the $x=0$
system. This subject is of importance for the understanding of the
system properties since (i) it is supposed to be crucial for the
superconducting phase~\cite{TheoEchSupra} and (ii) it is more
generally directly related to the systems conduction properties since
the conduction is supposed to occur in the ladder subsystem.

In such a compound, the localization of the electrons will mainly
arise from (i) the modulation of the on site energies due to
structural modulations (ii) the correlation between magnetic
electrons. A full treatment of this many body problem in an 
incommensurate system, is beyond the aim of this work, since it
focuses on the role of the structural distortions. Moreover, the
variations of the orbital energies are very large compared to the other
parameters of this model. In the following, we will thus consider only the
very first effect of electrons interactions, by including NN
repulsions. This approach will already give indications on the role of
the distortions on the electron localization, and transfer.

\begin{figure}[ht] 
\resizebox{8.5cm}{!}{\includegraphics{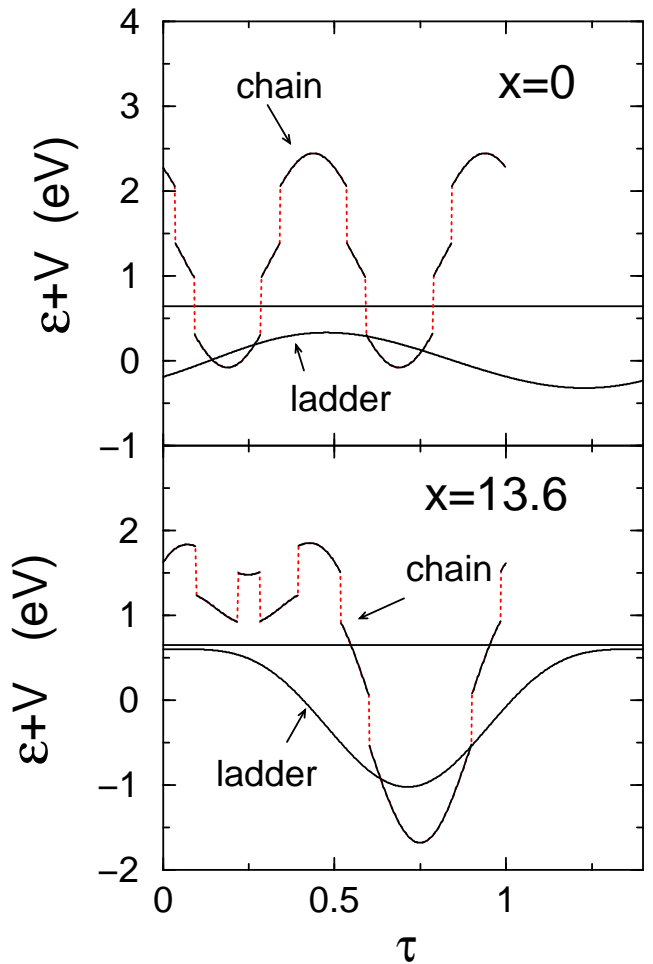}}
\caption{Chain and ladder energies as a function of
$\tau_{chains}$. The site energies were evaluated as the sum of the
orbital energies and the repulsion terms with neighboring atoms when 
the latter have lower orbital energies and are thus occupied prior
to the considered one. The ladder fourth coordinate $\tau_{ladder}$
has been rescaled to $2 c_{chain}/c_{ladder}$, so that the relative
$\tau$ variation range of the two subsystems is proportional to the
number of sites per f.u. in each sub-system. The horizontal  line
represents the Fermi level.}
\label{f:ece}
\end{figure}

We evaluated the relative energy of the chain and ladder subsystems,
responsible for the possible hole transfer, from the following
parameters~:
\begin{itemize}
\item the on-site orbital energies,  
\item the NN bi-electronic repulsion in the chain, the NN and NNN
bi-electronic repulsion in the ladder.
\end{itemize}
In references~\cite{sr14,ca13}, the NN bi-electronic repulsion $V$ was
computed for the chain subsystem using three centers fragment
calculations. It was found to be almost independent of the structural
modulations, temperature and calcium content. This can be understood
by the fact that the NN repulsions are essentially dependent on the
$C\!u$--$C\!u$ distances and not on the $C\!u$--$O$--$C\!u$ angles
that dominate the modulations of the other effective interactions.  We
thus evaluated the NN repulsions on the ladders subsystem using the
average computed chain value of $0.6\ \rm eV$ and the standard Ohno
formula~\cite{ohno} for the distance dependence
$$ V(R) = \frac{V_0}{1 + R/a_0}$$ where $R$ is the inter-site
distance, $a_0$ is the Bohr radius and $V_0$ a $R=0$ effective
constant.
Extracting $V_0$ from the chain repulsions we get $V_0=3.7\ \rm eV$. 
We thus find 
\begin{itemize}
\item for the intra-ladder repulsions~: $V_{leg} \simeq V_{rung}
\simeq 0.44\ \rm eV$, and $V_{diag} \simeq 0.32\ \rm eV$ (diagonal
repulsion on a plaquette),
\item for the inter-ladder repulsions~: $V_{leg-leg} \simeq 0.6\ \rm eV$. 
\end{itemize}
Using these approximations and the computed orbital energies we
determined the relative energies of the chains and ladders as a
function of $\tau$ (see figure~\ref{f:ece}). Since the ladder
hole-doping remains small we considered that the repulsions where
acting on all ladder sites. As a matter of comparison we computed the
mean (periodic approximation) relative energy between the ladder and
chain sites using the same method. We found for the $x=0$ undoped
compound the chains about $0.5eV$ above the ladders while for the
$x=13.6$ compound the chain and ladders subsystems are quasi-degenerate.
For the real compounds, one can see on figure~\ref{f:ece} that while
for the $x=0$ compound the ladder energy curve is always about $0.65\
\rm eV$ lower than the first empty sites of the chains, for the
$x=13.6$ compound, the most energetic ladder sites are at the Fermi
level. These results are coherent with a complete hole localization on
the chain sub-system for the undoped compound and a small hole
transfer to the ladders for highly doped systems.  It seems however
difficult to extract a precise value of the chain to ladder hole
transfer (for the $x=13.6$ compound). Indeed the hole transfer can be
expected to be sensitive to very small energetic variations, since the
Fermi level is located in a region with large density of states, both
for the chain and ladder subsystems.  For the ladders subsystem, $E_F$
is located on an energetic plateau. For the chain subsystem, $E_F$
crosses the energy curve, unlike what happens in the undoped system
where the Fermi level is located in an energetic gap. This may explain
the large experimental range of values obtained for this hole
transfer.

Another consequence of the Fermi level localization on a ladders
energetic plateau, is that the holes are localized on (quasi)
iso-energetic rungs ---~while they are localized on very energetically
different sites in the chain subsystem. These rungs are not first
neighbors, but in most cases second and in some cases third neighbors.
The holes can thus be considered as evolving on a subset of the
ladders with (quasi) iso-energetic sites.  In such a model, the rungs
belonging to the real ladder located in between two
hole-supporting rungs can be considered as bridging ligands. The
resulting effective hopping and exchange integrals between the
hole-supporting rungs are thus modulated according to the number and
the characteristics of these in-between rungs.  In such a model, these
effective interactions between hole-supporting rungs can be expected
to be rescaled to much lower values than the NN ones on the real
ladder, and thus much lower values than the on-rung ones ($|t_{\rm
leg}^{\rm eff}| \ll |t_{\rm rung}|$ and $|J_{\rm leg}^{\rm eff}| \ll
|J_{\rm rung}|$).

We would like to conclude this work on a few opened questions on the
super-conducting phase. First, the on-site energy variations are
quite-large on the calcium-doped system, both for the chains and
ladders subsystems. These modulations strongly increase as a function
of the chemical pressure. Whether the applied pressure would act along
the same line, it would induce an even larger distortion of the chains
and ladders subsystems. Such a phenomenon would be quite incompatible
with the usual super-conductivity theory in the ladder systems since
it would induce an even larger modulation of the ladder on-site
orbital energies. Another discrepancy between the present results and
the super-conductivity predictions is that the super-conductive
pairing is supposed to occur for (i) low ladder hole doping, (ii) $2 |J_{rung}|
> |J_{leg}|$ (both are verified in the present case) and (iii) $|J_{rung}| >
|t_{rung}|$ which is far from being verified since $|t_{rung}| > 4
|J_{rung}| $.  One thus wonder either what type of structural
distortions under applied pressure could correct these features
hindering the super-conductivity or whether the  observed
super-conductivity arise from another origin that the usually accepted
homogeneous ladder $t-J$ model.

{\bf Acknowledgments :} {\sf the authors thank Dr. J. Etrillard for
providing us with the undoped compound crystallographic structures,
Dr. Maynau for providing us with the CASDI code
. The present calculations were done at the CNRS/IDRIS computational
facilities under project n° 1104.}

\section*{Appendix}

The computed results were fitted as a function of
the fourth crystallographic coordinate $\tau$, using a Fourier series,
according to the following expression
\begin{eqnarray} \label{eq:fit}
&& a_0 + \sum_n a_n \cos{\left[2\pi n (\tau - \varphi_n) \right]}
\end{eqnarray}
Only terms with a non negligible contribution to the series were
retained.
 
The results are summarized in tables~\ref{t:fits}a for the $x=0$
and~\ref{t:fits}b for  $x=13.6$. 

\begin{table}[t]
(a) \begin{tabular}[t]{r|@{\hspace{1ex}}*{5}{r@{\hspace{3ex}}}r}
$x=0$ & \multicolumn{1}{c}{$\varepsilon_l$} &
\multicolumn{1}{c}{$\Delta \varepsilon_r$} & \multicolumn{1}{c}{$t_l$} &
\multicolumn{1}{c}{$t_r$} & \multicolumn{1}{c}{$J_l$} &
\multicolumn{1}{c}{$J_r$ } \\ \hline
$a_0$ &   0.0 & 0.0 & -659.2 & -572.2 & -143.9 & -111.6 \\
$a_1$ &   -153.0 & 204.3 & 28.3 & 12.4 & -13.7 & 2.5 \\
$a_2$ &   8.9 & 3.4 & 2.0 & -5.1 & -1.0 & -2.5 \\
\\               
$\varphi_1$ &   0.819 & 0.875 & 0.655 & 0.750 & 0.904 & 0.750 \\
$\varphi_2$ &   0.786 & 0.688 & 0.089 & 0.500 & 0.982 & 0.500 \\
\end{tabular}
(b) \begin{tabular}[t]{r|@{\hspace{1ex}}*{5}{r@{\hspace{3ex}}}r}
$x=13.6$ & \multicolumn{1}{c}{$\varepsilon_l$} &
\multicolumn{1}{c}{$\Delta \varepsilon_r$} & \multicolumn{1}{c}{$t_l$} &
\multicolumn{1}{c}{$t_r$} & \multicolumn{1}{c}{$J_l$} &
\multicolumn{1}{c}{$J_r$ } \\ \hline
$a_0$ &   0.0 & 0.0 & -672.5 & -624.7 & -159.3 & -136.4 \\
$a_1$ &   -811.4 &  & 73.9 & 11.5 & -0.2 & 8.0 \\
$a_2$ &   -209.5 &  & -11.1 & 4.0 & -10.5 & 4.5 \\
$a_3$ &    &  &  &  & 4.4 &  \\
\\               
$\varphi_1$ &   0.500 & 0.000 & 0.284 & 0.001 & 0.284 & 0.995 \\
$\varphi_2$ &   0.004 & 0.000 & 0.529 & 1.000 & 0.537 & 0.997 \\
$\varphi_3$ &    &  &  &  & 0.972 &  \\
\end{tabular}
\caption{Analytic fit of the $t-J$  model, a) for the
$x=0$ undoped compound, b) for the $x=13.6$ calcium-doped
compound. All energies are given in meV.}
\label{t:fits}
\end{table}

 \end{document}